\documentclass[amsmath,superscriptaddress,showpacs,prb,twocolumn]{revtex4-1}
\usepackage{bbm}
\usepackage{bm}
\usepackage{enumerate}
\usepackage{graphicx}
\usepackage[dvips]{epsfig}
\usepackage{epsf}
\usepackage[latin1]{inputenc}
\usepackage{amsmath}
\usepackage{amsfonts}
\usepackage{amssymb}
\usepackage{xcolor}
\usepackage{blindtext}

\usepackage{amsthm,bm, braket,color,comment,dcolumn}
\usepackage{hyperref,textcomp,gensymb,mathtools,url}
\usepackage[dvips]{epsfig}
\usepackage{overpic}

\bibpunct{[}{]}{;}{n}{}{}

\newcommand{\MS}{\text{MoS}_{2}}
\newcommand{\WS}{\text{WS}_{2}}
\newcommand{\WSe}{\text{WSe}_{2}}
\newcommand{\XMS}[1]{\text{{#1}/MoS}_{2}}
\newcommand{\XWS}[1]{\text{{#1}/WS}_{2}}
\newcommand{\XWSe}[1]{\text{{#1}/WSe}_{2}}
\DeclareMathOperator{\arctantwo}{Arg}

\begin{document}
	\title{Spirals and skyrmions in antiferromagnetic triangular lattices}
	
\author{Wuzhang Fang}
\author{Aldo Raeliarijaona}
\author{Po-Hao Chang}
\author{Alexey A. Kovalev}
\author{Kirill D. Belashchenko}
\affiliation{Department of Physics and Astronomy and Nebraska Center for Materials and Nanoscience, University of Nebraska-Lincoln, Lincoln, Nebraska 68588-0299, USA}
	
\date{\today}

\begin{abstract}
We study realizations of spirals and skyrmions in two-dimensional antiferromagnets with a triangular lattice on an inversion-symmetry-breaking substrate.
As a possible material realization, we investigate the adsorption of transition-metal atoms (Cr, Mn, Fe, or Co) on a monolayer of $\MS$, $\WS$, or $\WSe$ and obtain the exchange, anisotropy, and 
Dzyaloshinskii-Moriya interaction parameters using first-principles calculations. Using energy minimization and parallel-tempering Monte-Carlo simulations, we determine the 
magnetic 
phase diagrams for a wide range of 
interaction 
parameters. We find that skyrmion lattices 
can
appear even with weak Dzyaloshinskii-Moriya interactions, but their stability 
is
hindered by magnetic anisotropy. Our results suggest that $\XMS{Cr}$, $\XMS{Fe}$, and $\XWSe{Fe}$ interfaces can host spin spirals formed from the 120$\degree$ antiferromagnetic states. Our results further suggests  that for other interfaces, such as $\XMS{Fe}$, the Dzyaloshinskii-Moriya interaction is strong enough to drive the system into a
three-sublattice skyrmion lattice 
in the presence of experimentally feasible external magnetic field.  
\end{abstract}

\maketitle

\section{Introduction}

Antiferromagnetic spintronics~\cite{RevModPhys.90.015005} is an actively developing field with many potential applications as it explores materials with unique properties, i.e., materials associated with lack of stray fields and ultrafast dynamics in THz range~\cite{Olejnk2018}. Concepts of topology have influenced the development of condensed matter physics, e.g., as can be seen in realizations of magnetic skyrmions in ferromagnets~\cite{Rossler2006,Muhlbauer2009,Yu2010}. 
Realizations of skyrmions in collinear antiferromagnets (AFMs) are also being explored~\cite{PhysRevLett.116.147203,Zhang2016,PhysRevB.96.060406,PhysRevB.100.100408,PhysRevLett.125.257201}. Studies of three-sublattice systems generalize above ideas to noncollinear antiferromagnets and demonstrate possibilities for realizations of various skyrmion phases~\cite{Rosales,Leonov2015,PhysRevLett.122.187203,PhysRevX.4.011023,Gao2020}.
It has recently been demonstrated that the topological charge in AFMs can be also fractionalized~\cite{Gao2020,PhysRevLett.125.257201} which can be interpreted as a formation of merons in square crystals of vortices and antivortices~\cite{PhysRevB.91.224407,Utkan2016,Kovalev2018,Yu2018}. 

Materials engineering has been used for realizations of room temperature skyrmions in ferromagnets~\cite{MoreauLuchaire2016,Woo2016,Boulle2016}. 
In this work, we explore possibilities of integrating transition metal dichalcogenides (TMDs) with magnetic transition metals to realize noncollinear AFMs capable of hosting skyrmion lattices. Studies have demonstrated utility of TMDs for realizations of spin-valves~\cite{Wang2015,Wu2015} and TMD-based spintronic devices~\cite{Allain2015,Wang2015,Wu2015,Safeer2019}. 
Contact between metals and TMDs can alter the electronic properties of the system and affect the efficiency of TMD-based devices~\cite{Caciuc2018,Allain2015,Zhang2019}.
It is thus crucial to study various proximity effects associated with the metal/TMD interfaces~\cite{Zutic2019}.
Due to the layered nature of TMDs, capping them with magnetic atoms seems to be natural in order to bring about novel properties. In a recent study of Fe/Ir(111) and TMD interface, it has been found that transition-metal atoms in the TMD monolayer acquire a metallic character and spin imbalance due to its chemisorption on Fe/Ir~\cite{Caciuc2018}.
The large spin-orbit coupling (SOC) in monolayer TMDs~\cite{Latzke2015} combined with structural symmetry breaking in magnetic heterostructures can potentially lead to sizable Dzyaloshinskii-Moriya interaction (DMI)~\cite{DzyaloshinskyJPCS1958,MoriyaPR1960}. 
It is thus crucial to investigate magnetic interactions in TMD-based magnetic systems and determine such phenomenological parameters as the strength of exchange interaction, the strength of DMI, and the single-ion magnetic anisotropy. This will also determine the feasibility of realizing various skyrmion phases in TMD-based triangular AFMs.

A very strong DMI in a triangular AFM can lead to appearance of a three-sublattice skyrmion crystal~\cite{Rosales}; however, it is not clear whether such strong DMI can be realized in real materials. It has been suggested that a relatively large DMI, $D>0.2 J$ with $J$ being the exchange energy, is required for stabilization of skyrmions in triangular AFMs~\cite{PhysRevB.96.024404,Mohylna2020}, which puts a strong constraint on realizations of three-sublattice skyrmons. Numerically, it has been demostrated that skyrmion lattices in triangular AFM can be stabilized even for weak DMI, $D<0.2J$~\cite{Mohylna2021}. In real materials, the presence of SOC will likely also lead to the apearance of the single-ion magnetic anisotropy.
Despite the existing studies of three-sublattice skyrmions in triangular AFMs, the analysis of such skyrmions for the experimentaly feasible DMI and in the presence of single-ion magnetic anisotropy is still lacking. 

In this paper, we study realizations of magnetic spirals and skyrmions in triangular AFMs in the presence of DMI, comparable to or smaller than the critical DMI strength, $D_c=0.2 J$~\cite{PhysRevB.96.024404}, and include the effect of single-ion magnetic anisotropy. As a possible material realization, we consider one layer of magnetic transition metal X, such that X=\{Co, Cr, Fe, Mn\}, adsorbed onto a monolayer of transition metal dichalcogenides ($\MS$, $\WS$, or $\WSe$). From ab initio calculations, we evaluate parameters describing magnetic interactions, namely the exchange interaction, the uniaxial anisotropy, and the DMI, by computing the difference in total energy of 
magnetic configurations. We find that \{Co, Cr, Fe, Mn\}/MoS$_{2}$, $\XWS{Mn}$, and $\XWSe{Fe}$ can be described as AFMs with uniaxial anisotropy and DMI. Uisng Monte Carlo simulations, we study phase diagrams of triangular AFMs in the presence of DMI and magnetic anisotropy, and assess the feasibility of realizing magnetic spiral and skyrmion phases.
Our Monte Carlo simulations indicate that skyrmion lattices will appear even for relatively weak DMI below $D_c=0.2 J$, but their stability will be hindered by magnetic anisotropy.
We find that in the absence of magnetic field the magnetic ground state of $\XMS{Cr}$, $\XMS{Fe}$, and $\XWSe{Fe}$ is a spin spiral on three sublattices of the triangular lattice while the ground state of $\XMS{Mn}$, $\XWS{Mn}$, and $\XMS{Co}$ is the 120$\degree$ N\'eel type configuration. Our analysis further suggests that $\XMS{Fe}$ can potentially host a three-sublattice skyrmion crystal in the presence of experimentally feasible magnetic fields.

\section{Methods}\label{Sec:Method}

In this section, we briefly describe methods used in analyzing triangular AFMs.
The phenomenological parameters for several material candidates are established by employing first-principles calculations. The zero temperature phase diagrams are obtained by minimizing the energy for a wide range of phenomenological parameters. The finite temperature phase diagrams are obtained by employing the feedback-optimized parallel tempering Monte Carlo simulations. 
\subsection*{First-principles calculations}
First-principles calculations are performed using the Vienna \textit{ab initio} simulation package (VASP) \cite{VASP} and open source package for Material eXplorer (OpenMX) codes \cite{OMX}. In VASP calculations, we use the generalized gradient approximation~\cite{PBE} and projector augmented wave potentials~\cite{PAW1,PAW2}. A $27\times27\times1$ k-point mesh is used to sample the Brillouin zone. The dense mesh of k-point is necessary to converge to the correct magnetic state and to get accurate total energy. The plane-wave cutoff energy is 500 eV and energy convergence criteria are $10^{-5}$ eV for structure relaxations and $10^{-6}$ eV for total energy calculations. The structures are fully relaxed until the magnitude of the force on the individual atom is less than 0.01 eV/\AA. A thick vacuum layer of 17 \AA~along the out-of-plane direction is used to minimize the interaction due to the periodic condition. To describe the interactions between localized d electrons of transition atoms accurately, we take into account the on-site Coulomb interactions using the LDA+U method implemented in VASP \cite{Uvasp} and OpenMX \cite{Uomx}. The effective on-site Coulomb parameters $U$ is calculated for each transition metal atom using the linear response method, as implemented in VASP, and the effective exchange parameter $J$ is chosen as 0.9 eV. After determining the adsorption sites and relaxed structures using VASP, we use OpenMX for the calculations of physical properties of our systems. OpenMX uses pseudo-atomic localized orbitals \cite{OMX_oribitals} as basis functions. Band structures have been compared to the ones calculated using VASP to determine the basis sets used in our calculations. 

\subsection*{Model Hamiltonian}

\begin{figure}
\includegraphics[width=0.85\columnwidth]{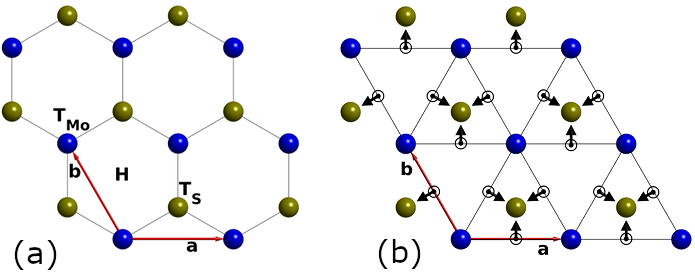}
\caption{(a) Three adsorption sites (H, T$_\text{S}$ and T$_\text{Mo}$) on top of monolayer $\MS$. Mo and S atoms are represented by blue and orange spheres, respectively; (b) the same unit cell with the Dzyaloshinskii-Moriya vector depicted by the black arrows centered on the bond of neighboring, interacting atoms. The convention taken for the direction on bonds is counterclockwise in a triangular plaquette with Mo atom at its center. The vectors \textbf{a} and \textbf{b} are the primitive vectors of the hexagonal Bravais lattice.}
\label{fig1}
\end{figure}

The monolayer of adsorbed transition atoms on top of $\MS$, $\WS$ or $\WSe$ form a triangular lattice. The neighboring triangles are not equivalent due to the existence of the transition metal
, and this asymmetry results in the Dzyaloshinskii-Moriya interactions (DMI) \cite{Dzyaloshinskii,Moriya}. The model Hamiltonian to describe the magnetic interactions between the adsorbed transition atoms is given by:
\begin{eqnarray}\label{Ham}
H&=&\sum_{\braket{ij}}{J\, \mathbf{S}_{i} \cdot \mathbf{S}_{j}  +\Tilde{\mathbf{D}}_{ij}\cdot(\mathbf{S}_i\times\mathbf{S}_j)}\nonumber \\
&-&\Tilde{K}\sum_{i}{(S_i^z)^2} -\Tilde{h}\sum_{i}{S_i^z},
\end{eqnarray}
where $i$ and $j$ are site indices with $\braket{ij}$ indicating the nearest neighbors; for each site we introduce a spin variable $\mathbf{S}_i$ with $|\mathbf{S}_i|=1$, $J$ describes the exchange interaction, $\Tilde{K}$ is the single-ion anisotropy constant, $\Tilde{h}=\mu B$ describes Zeeman coupling of the magnetic moment $\mu$ with the external magnetic field $B$, and $\Tilde{\mathbf{D}}_{ij}=\Tilde{\mathbf{D}}_{\perp} + \Tilde{\mathbf{D}}_{\parallel}$ is the DMI vectors, as shown in Fig.~\ref{fig1}, split into out-of-plane, $\Tilde{\mathbf{D}}_{\perp}$, and in-plane, $\Tilde{\mathbf{D}}_{\parallel}$, contributions.
We introduce the reduced parameters $\mathbf{D}_{ij}=\Tilde{\mathbf{D}}_{ij}/J$, $h=\Tilde{h}/J$, $K=\Tilde{K}/J$, and $T=k_B\Tilde{T}/J$, which we will use except when we refer to parameters determined from first-principles. 

\subsection*{Energy minimization}

To find the classical ground state in the absence of magnetic field at zero temperature, we use energy minimization.
This can be achieved by using the spherical approximation within the Luttinger-Tisza method~\cite{LuttingerTisza}. We replace the constraint $|\mathbf S_{i}|=S$ by the milder constraint $\sum_i |S_{i}|^2=N S^2$ where $N$ is the number of lattice sites. As an alternative approach, we also directly substitute a spiral ansatz in Eq.~\eqref{Ham} and perform numerical minimization, which slightly improves on the spherical approximation.
The model Hamiltonian in Eq.~\eqref{Ham} is rewritten using the Fourier transform (see Appendix~A) of the spins ${\mathbf{ S}_{\mathbf{ r},a} =\sum_{\mathbf{ q}}\mathbf{ S}_{\mathbf{ q},a} e^{i \mathbf{ q \cdot r}}}$ as:
\begin{equation}\label{Hamq}
H = \sum_{\mathbf{ q}}{\mathbf{ \Psi}_{\mathbf{ -q}} H_\mathbf{ q} \mathbf{ \Psi}_{\mathbf{ q}}},
\end{equation}
where $H_\mathbf{ q}$ is the Fourier transformed Hamiltonian, and the vector $\mathbf{  \Psi_{q} }$ is given by:
\begin{equation}
\mathbf{  \Psi_{q} } =\{ \mathbf{ S}_{\mathbf{ q},1},\mathbf{ S}_{\mathbf{ q},2},\mathbf{ S}_{\mathbf{ q},3} \}.
\end{equation}
Within the Luttinger-Tisza formalism, the ground state is determined by solving the eigenvalue problem $H_{\mathbf{q}} \Psi_{\mathbf{q}} = \varepsilon_{\mathbf{q}} \Psi_{\mathbf{q}}$ where the constraint leads to the normalization of the eigenvectors~\cite{LuttingerTisza}. To improve this procedure, we also numerically minimize Eq.~\eqref{Hamq} by directly substituting a spiral magnetic structure. 

\subsection*{Parallel tempering Monte Carlo}

To obtain the phase diagram and perform simulated annealing, we use feedback-optimized parallel tempering Monte Carlo simulations with Metropolis updates~\cite{PhysRevLett.57.2607,Hukushima1996}. We used up to 300 spin configurations (replicas), simultaneously simulated at different temperatures. The temperatures are chosen according to optimization algorithm improving the exchanges of states of different replicas~\cite{Katzgraber2006}, see Appendix B for details. The system consists of $N = L^{2}$ sites for $L = 75$ with periodic boundaries (in some cases we considered larger systems increasing the systems size to $L=150$). The system was equilibrated using runs of $10^6-10^7$ Monte-Carlo sweeps (attempts per spin) and consequently the same number of sweeps was used to accumulate statistics. By considering different lattice sizes, we observe finite size effects; however, we observe that our results capture all important features of the phase diagram given the large size of the system.
The magnetic ground state is characterized using the real space spin texture in combination with the spin structure factor defined as:
\begin{gather}\label{Sqperp}
    S_{\perp}(\mathbf{q}) = \frac{1}{N}\bigg\langle\sum_{\alpha=x,y}{ \left| \sum_{\mathbf{r} }{S^{\alpha}_{\mathbf{r}} \exp^{- i \mathbf{q}\cdot\mathbf{r}}} \right|^{2} }\bigg\rangle, \\
    \label{Sqpara}
    S_{\parallel}(\mathbf{q}) = \frac{1}{N}\bigg\langle \left| \sum_{\mathbf{r} }{S^{z}_{\mathbf{r}} \exp^{-i \mathbf{q}\cdot\mathbf{r}}} \right|^{2}  \bigg\rangle,
\end{gather}
where $\langle\rangle$ stands for statistical averaging, $N$ is the number of spins, and the perpendicular or parallel components to the plane normal are considered. In addition, we introduce the sublattice topological charge~\cite{Berg1981} by summing contributions from all elementary triangles of sublattice $\alpha$: 
\begin{equation}\label{Qtop}
    Q_\alpha=\frac{1}{2 \pi} \sum_\Delta \arctantwo(A_{\alpha,\Delta}+i B_{\alpha,\Delta}),
\end{equation}
where
\begin{eqnarray}
    A_{\alpha,\Delta}&=&\mathbf{S}_{\alpha,1}\cdot\left(\mathbf{S}_{\alpha,2} \times \mathbf{S}_{\alpha,3}\right),\\
    B_{\alpha,\Delta}&=&1+\mathbf{S}_{\alpha,1}\cdot\mathbf{S}_{\alpha,2}+\mathbf{S}_{\alpha,2}\cdot\mathbf{S}_{\alpha,3}+\mathbf{S}_{\alpha,1}\cdot\mathbf{S}_{\alpha,3},
\end{eqnarray}
with $\mathbf{S}_{\alpha,1}$, $\mathbf{S}_{\alpha,2}$, and $\mathbf{S}_{\alpha,3}$ being spins located on vertices of elementary triangle of lattice $\alpha$. The total topological charge over all three sublattices is defined as $Q=\sum_{\alpha=1}^3Q_\alpha$. 
To identify the phase boundaries, we calculate the specific heat:
\begin{gather}\label{Cv}
C_{\text{V}} = \frac{1}{T^2}\left(\braket{E^{2}}-\braket{E}^{2}\right),
\end{gather}
and the topological susceptibility:
\begin{equation}\label{chi}
    \chi_{{\pm}}=\frac{\langle (Q_{\pm})^2\rangle-\langle Q_{\pm}\rangle^2}{\langle Q_{\pm}\rangle T},
\end{equation}
where we define skyrmion-like, $Q_+$, and antiskyrmion-like, $Q_-$, topological charge~\cite{PhysRevB.39.7212,Bttcher2018}, k$_\text{B}$ is the Boltzmann constant, and $T$ is the reduced temperature. Note that only triangles with positive charge density contribute to $Q_{+}$ and only triangles with negative charge density contribute to $Q_{-}$, so that $Q=Q_{+}-Q_{-}$.
In the presentation of our results, we use $\chi_{-}$ as the $\chi_{-}$ peak is higher than the $\chi_{+}$ peak. According to our results, the positions of both peaks are almost identical.  

\begin{figure}
\includegraphics[width=1\linewidth]{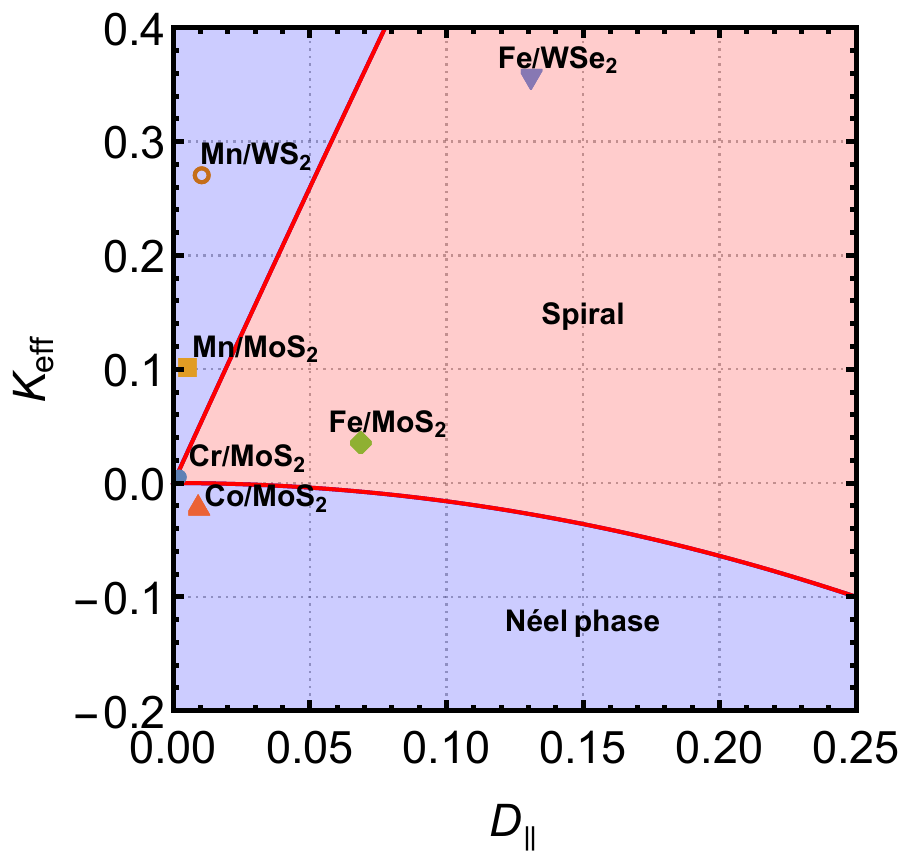}
\caption{Phase diagram as a function of DMI and effective anisotropy of AFM triangular lattice at zero temperature in the absence of magnetic field. The horizontal axis indicates the in-plane DMI and the vertical axis indicates the effective uniaxial anisotropy.
The plot shows the ground states of (Cr, Fe, Co, $\XMS{Mn)}$, $\XWS{Mn}$ and $\XWSe{Fe}$. The colors mark the different phases, i.e., the light blue region corresponds to the N\'eel phase and the pink region is the spiral phase. The plotting range was chosen to distinctly display all 6 cases in Table~\ref{table}.}
 \label{fig3}
\end{figure} 

\begin{figure}
 \includegraphics[width=0.8\linewidth]{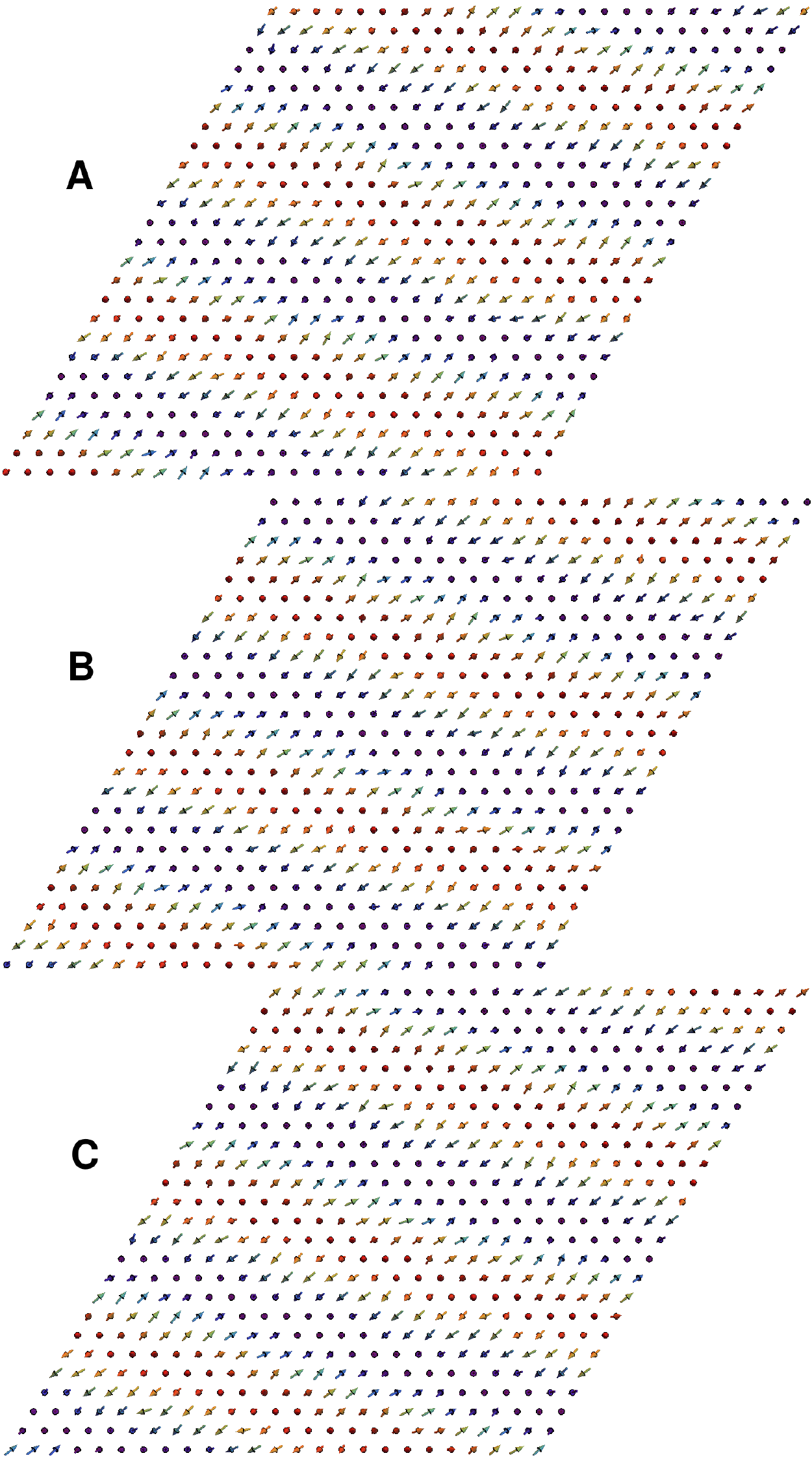}
 \caption{A snapshot of spin spiral in a triangular AFM obtained by simulated annealing for $K_\text{eff} = 0.4$, $D_{\parallel} = 0.2 $, $h = 0$, and $T=0.02$ on a $75\times 75$ triangular lattice with periodic boundary conditions. The letters on the left side (\textbf{A, B, C}) indicate the plotted sublattice. The colors indicate the orthogonal to the plane component of the spins.}
 \label{fig4}
\end{figure} 

\section{Results and discussion}\label{Sec:Results}

In this section, we first discuss results of first-principles calculations establishing adsoption sites of transition metal atoms on top of the monolayer TMD. These calculations further establish phenomenological parameters describing magnetic interactions for several material candidates. Finally, we discuss zero and finite temperature phase diagrams for a wide range of phenomenological parameters describing triangular AFMs. Particular emphasis is given to realizations of SkX in the presence of weak DMI and magentic anisotropy.

\subsection*{Adsorption sites, electronic structure, and magnetic ordering}

As shown in Fig.~\ref{fig1}, three possible adsorption sites of transition-metal atoms on monolayer TMD are denoted as H (hollow site), T$_\text{S}$ (on top of S), and T$_\text{TM}$ (on top of TM, where TM is either Mo or W).
We consider the case where all adsorption sites of a given type are occupied. 
The calculated $U$ values and preferred adsorption sites are listed in Table~\ref{table}.

\begin{table*}
\caption{Parameters in the model Hamiltonian.}
\centering
\begin{tabular}{ | c | c | c | c | c | c | c | c | c | c |} 
\hline
Atom & Adsorption site & $U$ (eV) & $J$ (meV) & T$_\text{N}$ (K) & $\Tilde{
\text{D}}_{\perp}$ (meV) & $\Tilde{\text{D}}_{\parallel}$ (meV) & $K$ (meV) & $K_\text{eff}$ (meV)\\
\hline
Cr/MoS$_2$ & On top of S & 4.9 & 37.04 & 430 & 0.003 & 0.08 & 0.18 & 0.194\\ 
\hline
Mn/MoS$_2$ & On top of S & 4.4 & 19.26 & 224 & 0.39  & 0.10 & -0.05 & 1.952\\ 
\hline
Fe/MoS$_2$ & Hollow   & 4.4 & 15.30 & 178 & -0.21 & 1.05 & 1.65 & 0.539\\ 
\hline
Co/MoS$_2$ & On top of S & 4.9 &  3.30 & 38  & 0.008 & 0.03 & -0.11 & -0.066\\
\hline
Fe/WSe$_2$ & Hollow & 4.4 & 12.45 & 144 & 0.68 & 1.63 & 0.86 & 4.42\\
\hline
Mn/WS$_2$ & On top of S & 4.4 & 23.04 & 267 & 1.03 & 0.24 & -0.88 & 6.23\\
\hline
\end{tabular}
\label{table}
\end{table*}

\begin{figure}
\includegraphics[width=1.0\columnwidth]{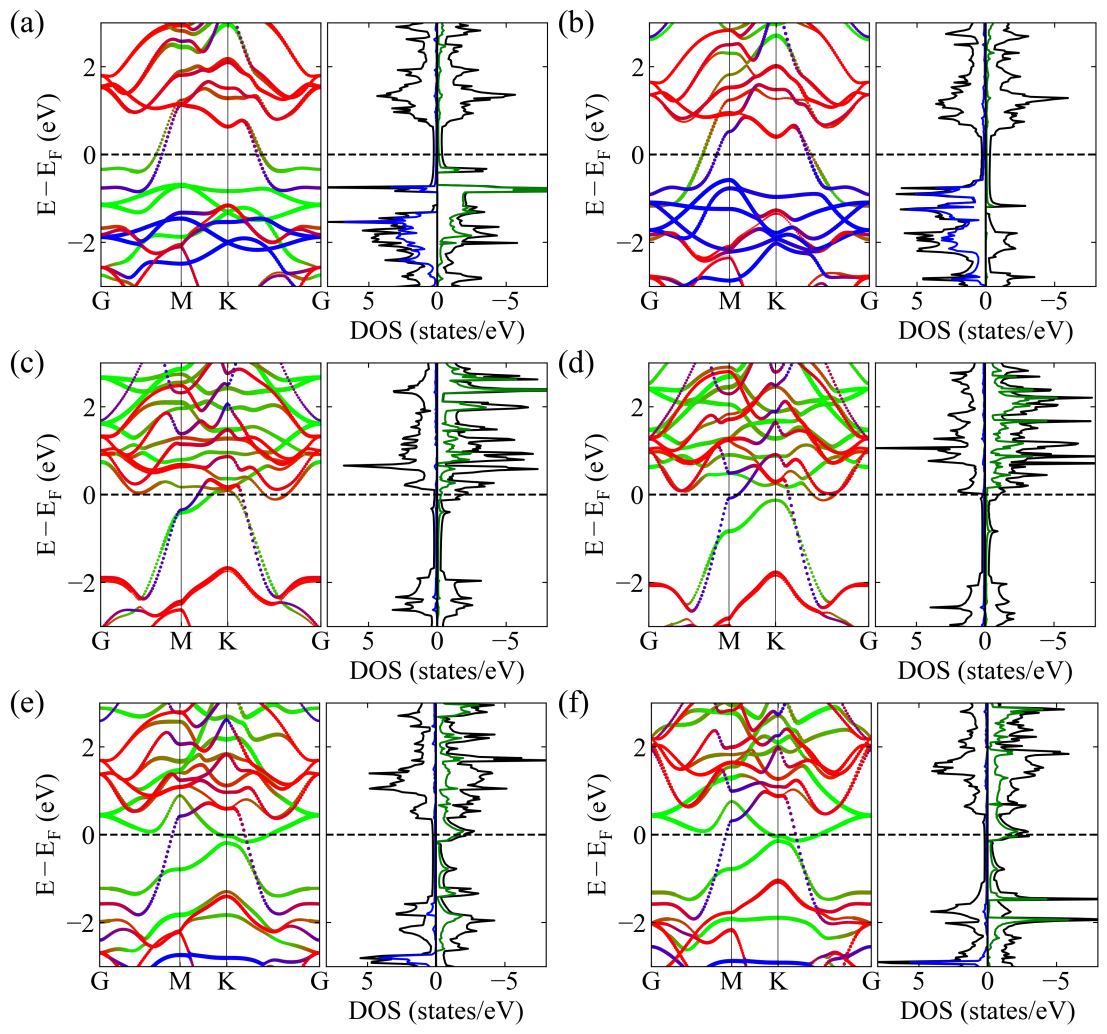}
\caption{Projected band structures and densities of states of (a) Co/MoS$_2$ (b) Cr/MoS$_2$ (c) Mn/MoS$_2$ (d) Mn/WS$_2$ (e) Fe/MoS$_2$ (f) Fe/WSe$_2$. The red, blue, and green colors in the band-structure plots represent the partial weights of Mo or W states, the majority-spin states of the magnetic atom, and the minority-spin states of the magnetic atom, respectively. Black lines in the DOS plots show the total DOS; blue and green, respectively, majority-spin and minority-spin partial DOS for the magnetic atom.}
 \label{bandstructure}
\end{figure}

The band structures and densities of states (DOS) in the ferromagnetic state for each system are shown in Fig.~\ref{bandstructure}. Red color represents Mo or W character; blue and green depict majority and minority-spin states of the magnetic transition-metal atom. In Co/MoS$_2$ and Cr/MoS$_2$, the bands crossing the Fermi level have a mixed character deriving from the hybridization of the conduction-band states of monolayer MoS$_2$ with the states of Co or Cr. These bands have a strongly $k$-dependent exchange splitting reflecting the varying weight of the Co or Cr states. In systems with Mn the Fermi level lies near the bottom of additional TMD-derived bands, while in systems with Fe there are almost pure minority-spin Fe bands at the Fermi level.

Next, we determine the preferred magnetic ordering of the adsorbed atoms on the adsorption sites. Ferromagnetic and $120\degree$ in-plane antiferromagnetic orderings are considered in our calculations. Total energies calculated with these two orderings suggest that the $120\degree$ in-plane antiferromagnetic ordering is the preferred magnetic state with the lowest total energy.

To extract the parameters of the model Hamiltonian \eqref{Ham}, we perform constrained DFT calculations with suitably specified non-collinear spin configurations, which are
described in Appendix C. The resulting parameters are listed in Table~\ref{table}.

\subsection*{Spiral classical ground state at zero temperature}

\begin{figure}
\includegraphics[width=0.49\columnwidth]{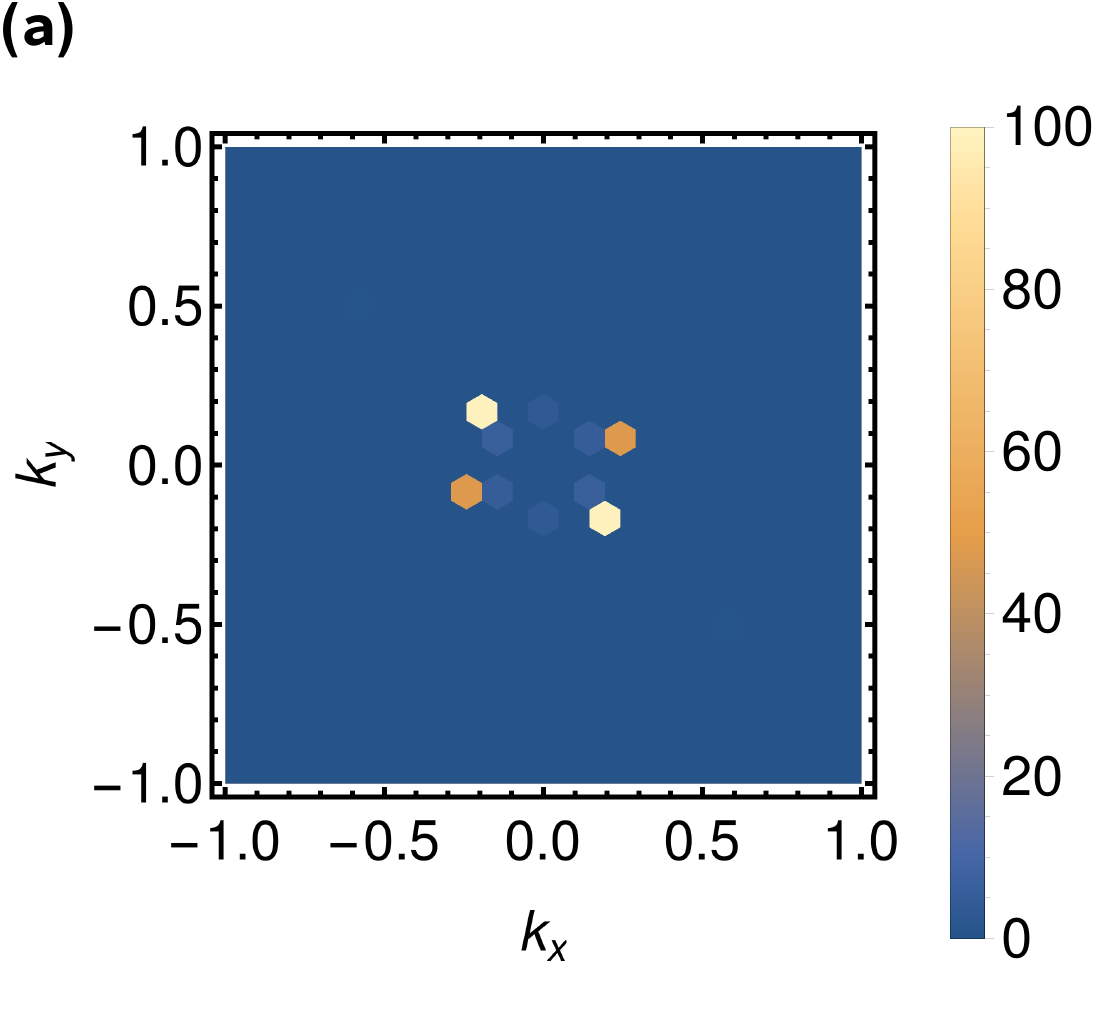}
\includegraphics[width=0.49\columnwidth]{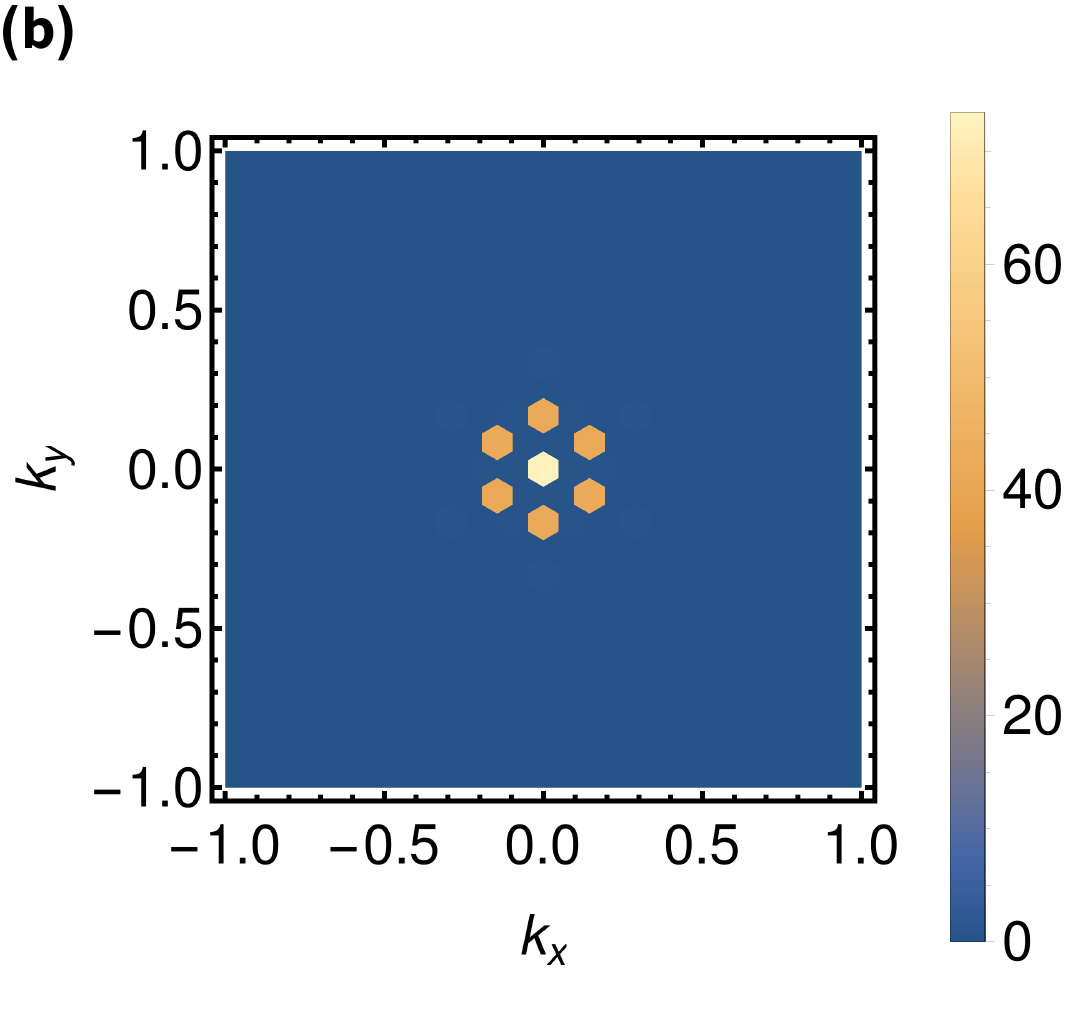}
\caption{Structure factor $S_{\parallel}(\mathbf{q})$ of a triangular AFM from Monte Carlo simulations. The vertical and horizontal axes represent the reciprocal space coordinates $k_\text{x}$ and $k_\text{y}$ in units of $1/L$.
The plot is for sublattice \textbf{A}; those for \textbf{B} and \textbf{C} are almost identical. (a) The plot corresponds to spiral phase realizable for $K = 0.4$, $D_{\parallel} = 0.2$, $h = 0$, and $T=0.02$. (b) The plot corresponds to SkX phase realizable for $K = 0$, $D_{\parallel} = 0.2$, $h = 3.1$, and $T=0.02$.}
 \label{fig5}
\end{figure}

In the absence of DMI and external magnetic field, the ground state is  AFM with a 120$\degree$ three-sublattice magnetic structure~\cite{Gvozdikova,Seabra,Rosales}. 
The spins in each sublattice separately are aligned and the angle between magnetizations from each sublattice is 120$\degree$.
In the presence of DMI, Eq.~\eqref{Hamq} has minimum at nonzero wave vector $\mathbf{k}_\text{min}$.
The ground state now contains three spirals with wave vector $\mathbf{k}_\text{min}$ set on each sublattice of the triangular lattice. 
It is convenient to introduce the effective anisotropy K$_\text{eff}$ that combines the effect of the uniaxial anisotropy $K$ and D$_{\perp}$ in Eq.~\eqref{Ham}, where K$_\text{eff} > 0$ prefers the easy axis alignment, whereas K$_\text{eff} <  0$ prefers the easy plane alignment.
By considering the energy difference between the out-of-plane and in-plane coplanar 120$\degree$ configurations, we can derive the expression of $K_\text{eff}$ to be:
\begin{equation}\label{Keff}
K_\text{eff} = K + D_{\perp} (3\sqrt{3}).
\end{equation}
The phase diagram resulting from energy minimization in Eq.~\eqref{Hamq} is depicted in Fig.~\ref{fig3}. 
By comparing energies for different magnetic structures, we distinguish two types of ground states: the three-sublattice N\'eel states (marked as the light blue region in Fig.~\ref{fig3}) in which the spins on the same sublattice are collinear and a spiral state (marked as the pink region in Fig.~\ref{fig3}).
To confirm the phase diagram in Fig.~\ref{fig3}, we also performed simulated annealing calculations. The resulting ground states have been characterized by observing the real space spin textures and the spin structure factors in the reciprocal space in Figs.~\ref{fig4} and \ref{fig5}.

The phase diagram in Fig.~\ref{fig3} 
shows that for 3 of the 6 compounds in Table~\ref{table}, $\XMS{Cr}$, $\XMS{Fe}$, and $\XWSe{Fe}$, the magnetic ground state is a spin spiral.
As an example, in Figs.~\ref{fig4} and \ref{fig5}(a) we show the real space spin textures and the spin structure factors for the case $K_\text{eff} = 0.4$, $D_{\parallel} = 0.2$.
Note that additional (substantially smaller in height) peaks in Fig.~\ref{fig5}(a) arise due to the fact that the spin structure factor is calculated by performing the statistical averaging at low but finite temperature, $T=0.02$. The dominant peak clearly identifies the spiral phase in Fig.~\ref{fig5}(a).

\begin{figure}
\includegraphics[width=0.49\columnwidth]{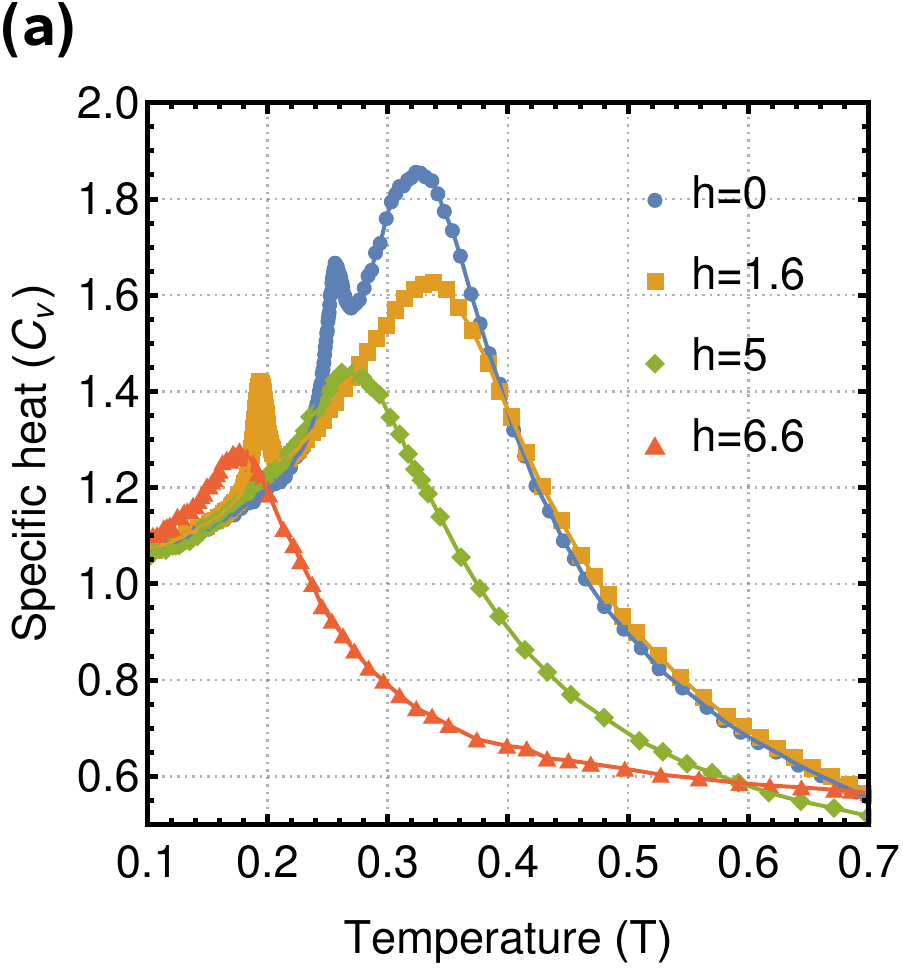}
\includegraphics[width=0.49\columnwidth]{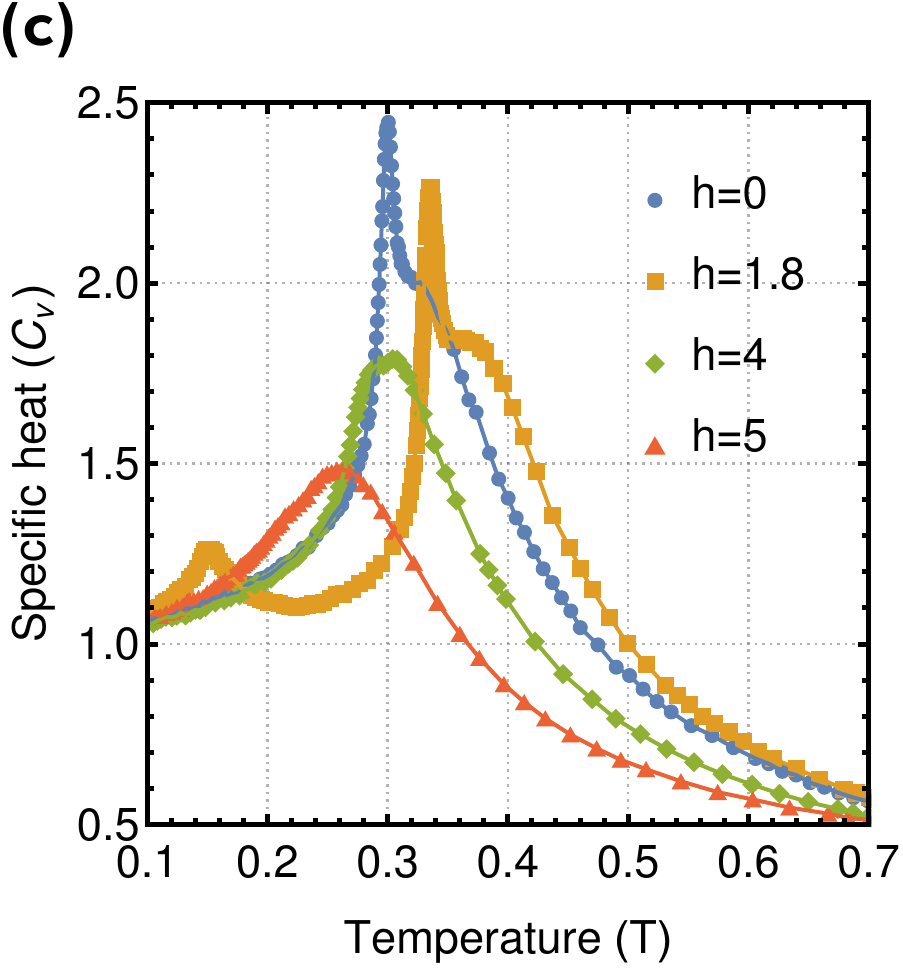}
\includegraphics[width=0.49\columnwidth]{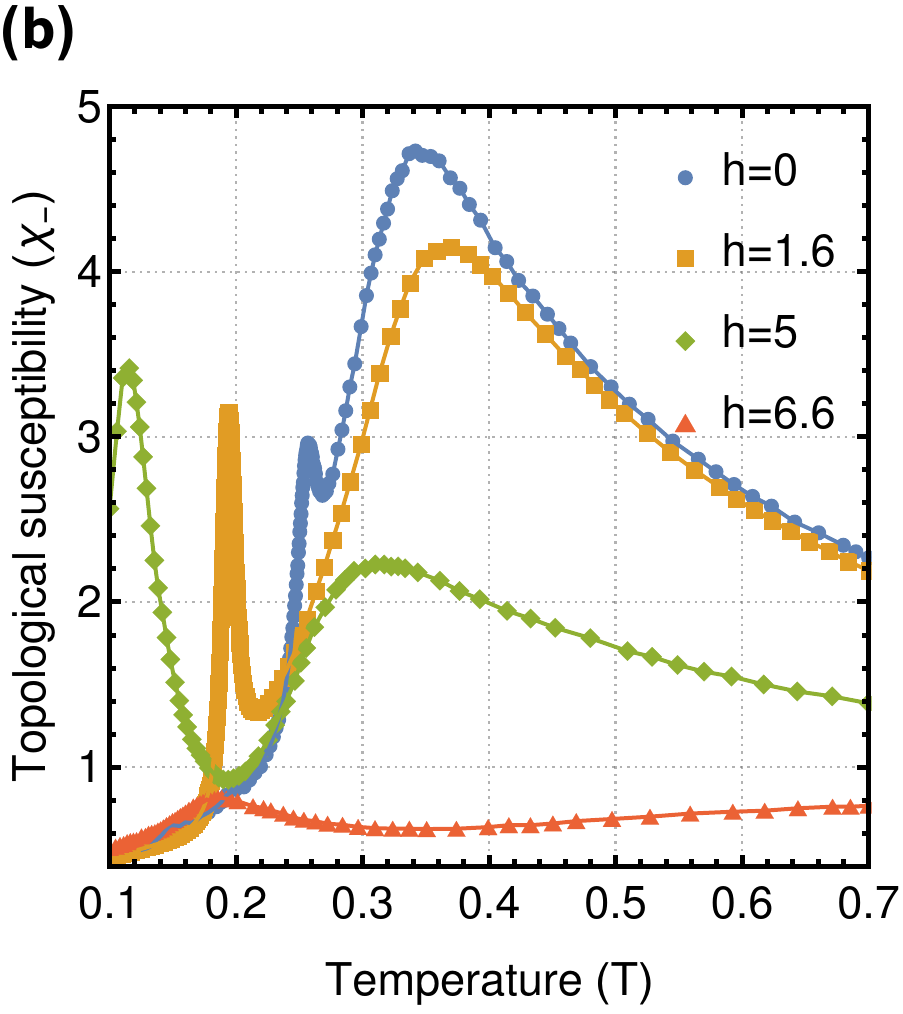}
\includegraphics[width=0.49\columnwidth]{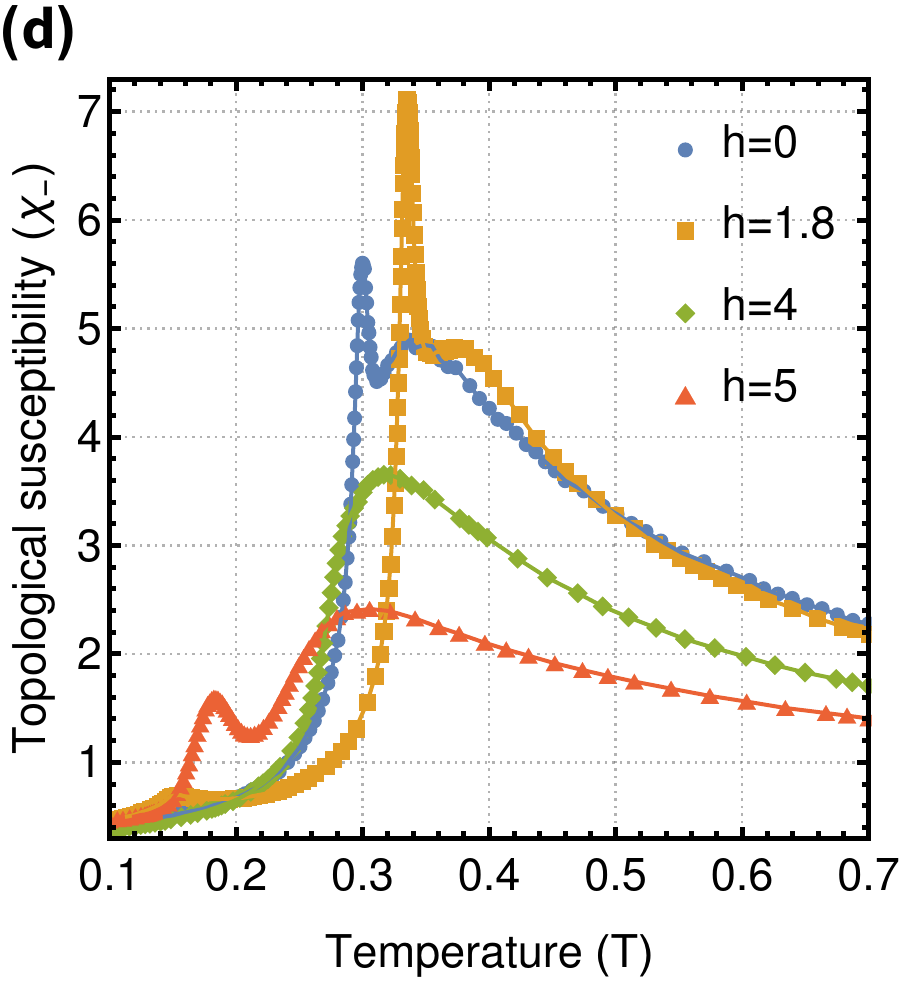}
 \caption{(a) and (b) Specific heat and topological susceptibility as a function of temperature for different values of the external magnetic field, $D_\parallel= 0.2$, $K = 0$. (c) and (d) Specific heat and topological susceptibility as a function of temperature for different values of the external magnetic field, $D_\parallel= 0.2$, $K = 0.1$.}
 \label{fig7}
\end{figure}

\subsection*{Antiferromagnetic skyrmion crystals}

In this section, we show that it is possible to stabilize antiferromagnetic skyrmion crystals (SkXs) at a finite temperature and in the presence of a magnetic field. To this end, we consider a $75 \times 75$ triangular lattice with periodic boundary conditions and perform parallel tempering Monte Carlo simulations. To identify the phase transitions, we calculate the specific heat, the topological susceptibility, and the total topological charge according to Eqs.~\eqref{Cv}, \eqref{chi}, and \eqref{Qtop}, respectively. Some of the curves used for identifying phase transitions are shown in Fig.~\ref{fig7} where we plot the specific heat and topological susceptibility as a function of temperature. In Figs.~\ref{fig7}(a) and (c), we plot the specific heat and topological susceptibility for $D_\parallel=0.2$, $K=0$ and for 4 values of the external field, $h=0,\,1.6,\,5,\,6.6$. In Figs.~\ref{fig7}(b) and (d), we plot the specific heat and topological susceptibility for $D_\parallel=0.2$, $K=0.1$ and for 4 values of the external field, $h=0,\,1.8,\,4,\,5$. The locations of phase transitions are identified by tracking the peaks in the specific heat and topological susceptibility. It is necessary to account for the finite size effects and scaling relations to pinpoint the exact locations of phase transitions. Even though we do not perform the scaling analysis, by comparing different system sizes we conclude that the chosen system size is large enough and allows us to correctly describe the schematic of the phase diagrams for our systems.

\begin{figure}
 \includegraphics[width=0.8\linewidth]{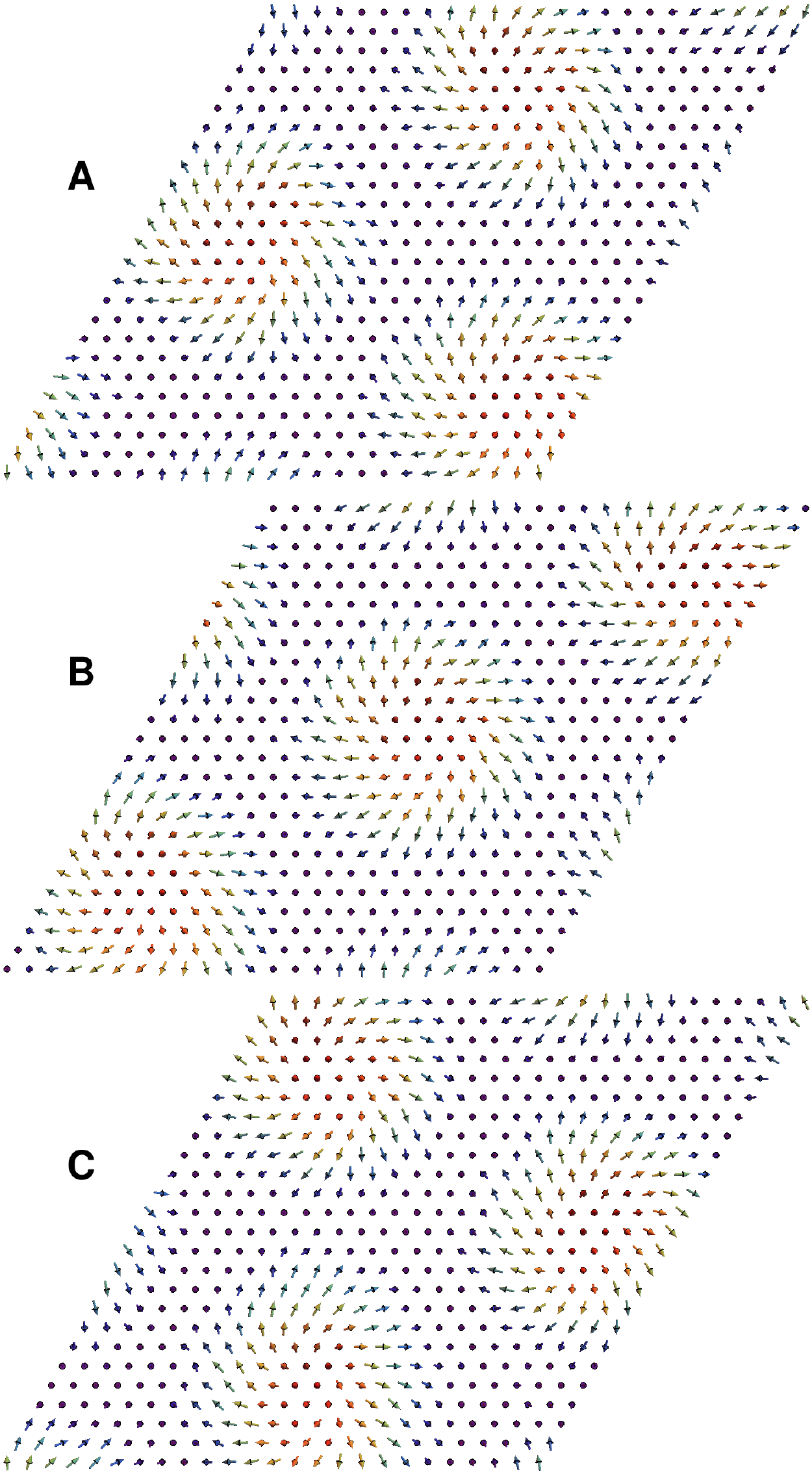}
 \caption{A snapshot of SkX in a triangular AFM obtained by simulated annealing for $K = 0$, $D_{\parallel} = 0.2 $, $h = 3.1$, and $T=0.02$ on a $75\times 75$ triangular lattice with periodic boundary conditions. The letters on the left side (\textbf{A, B, C}) indicate the plotted sublattice. The colors indicate the orthogonal to the plane component of the spins.}
 \label{fig8}
\end{figure} 

\begin{figure*}
\includegraphics[width=0.9\columnwidth]{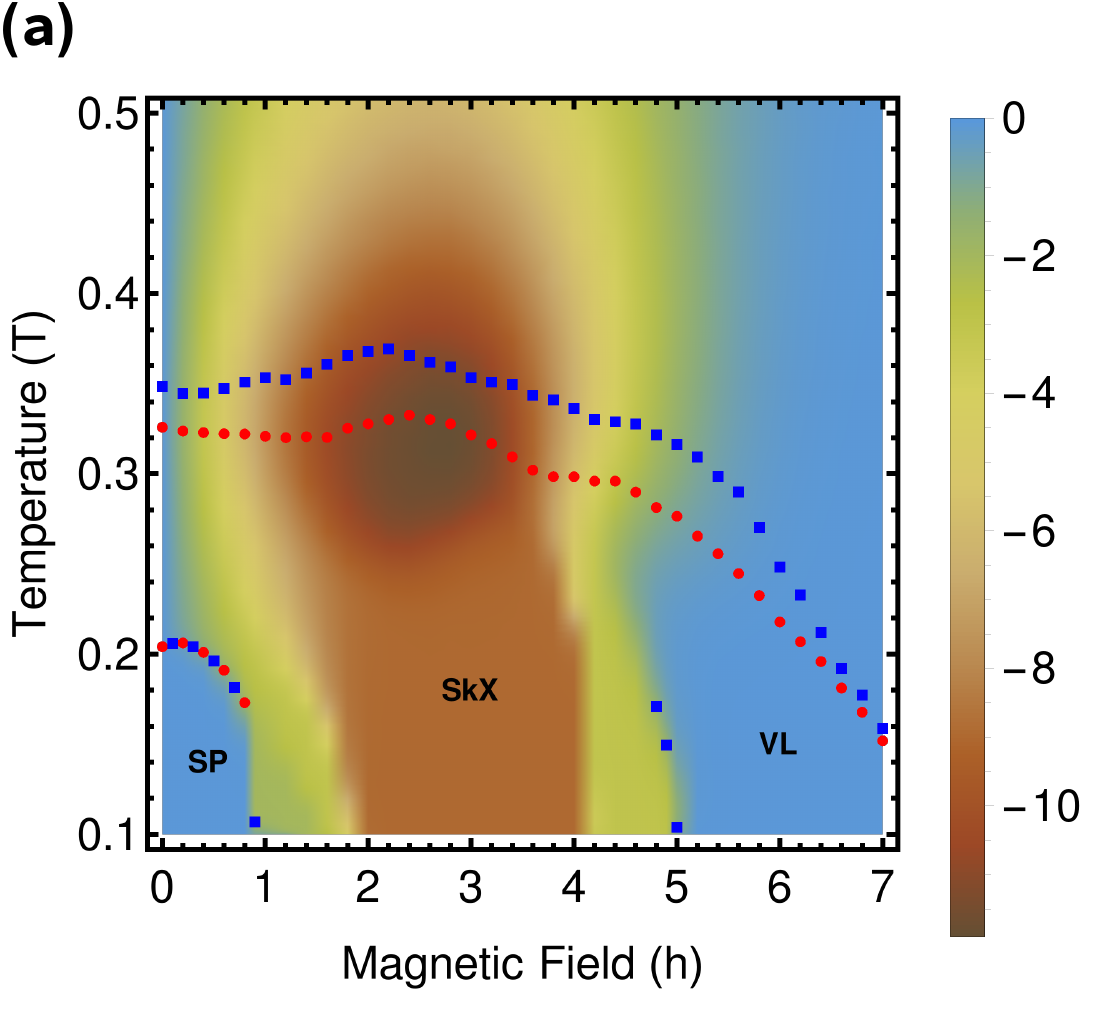}
\includegraphics[width=0.9\columnwidth]{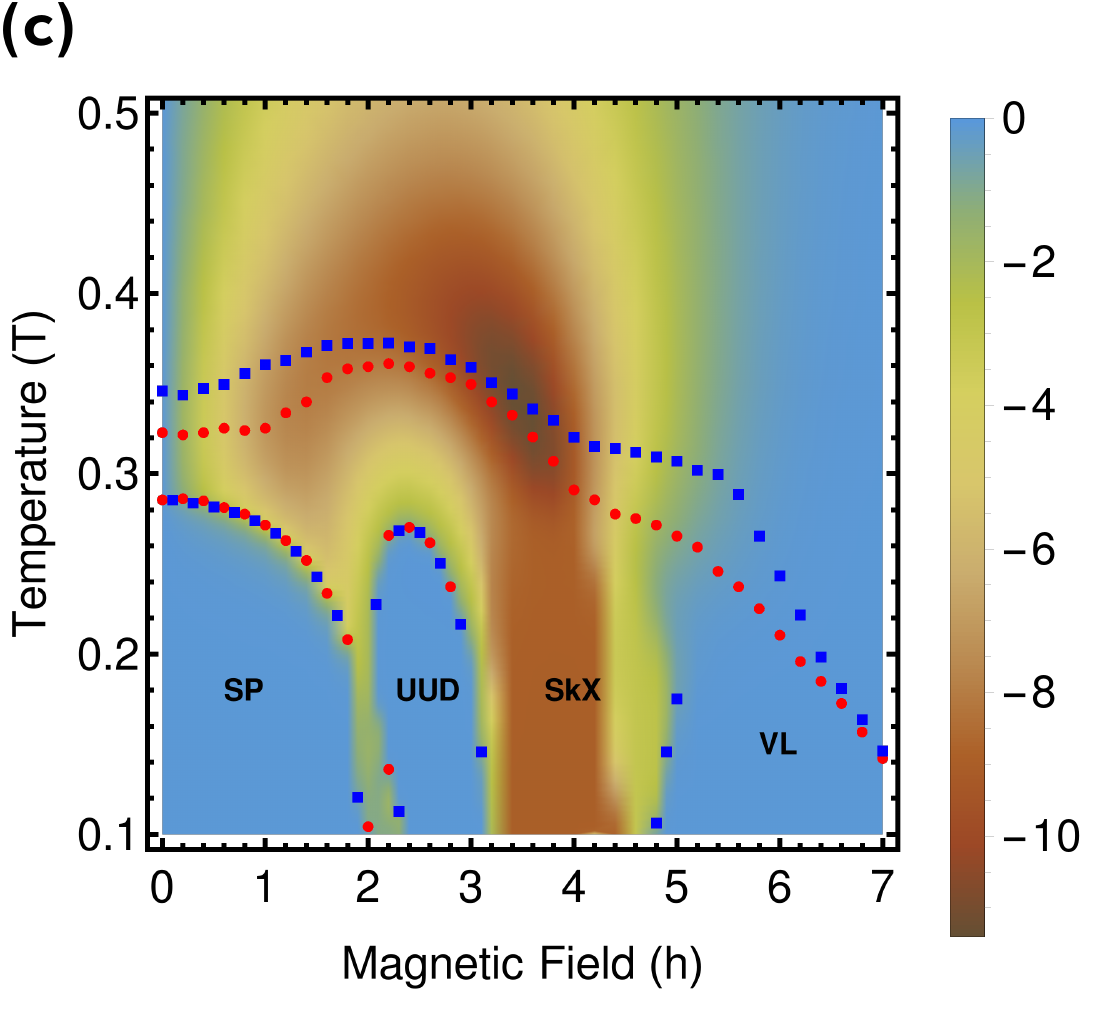}
\includegraphics[width=0.9\columnwidth]{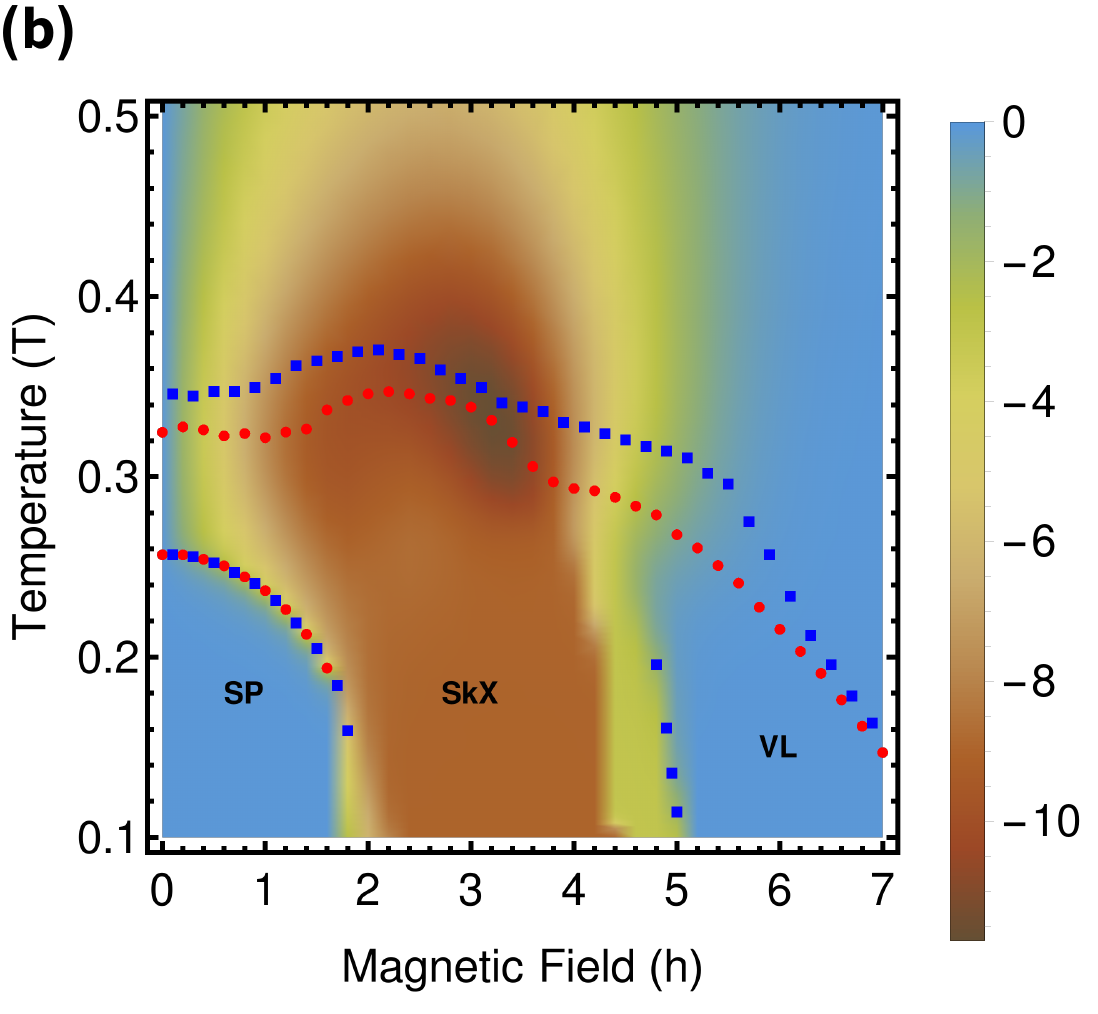}
\includegraphics[width=0.9\columnwidth]{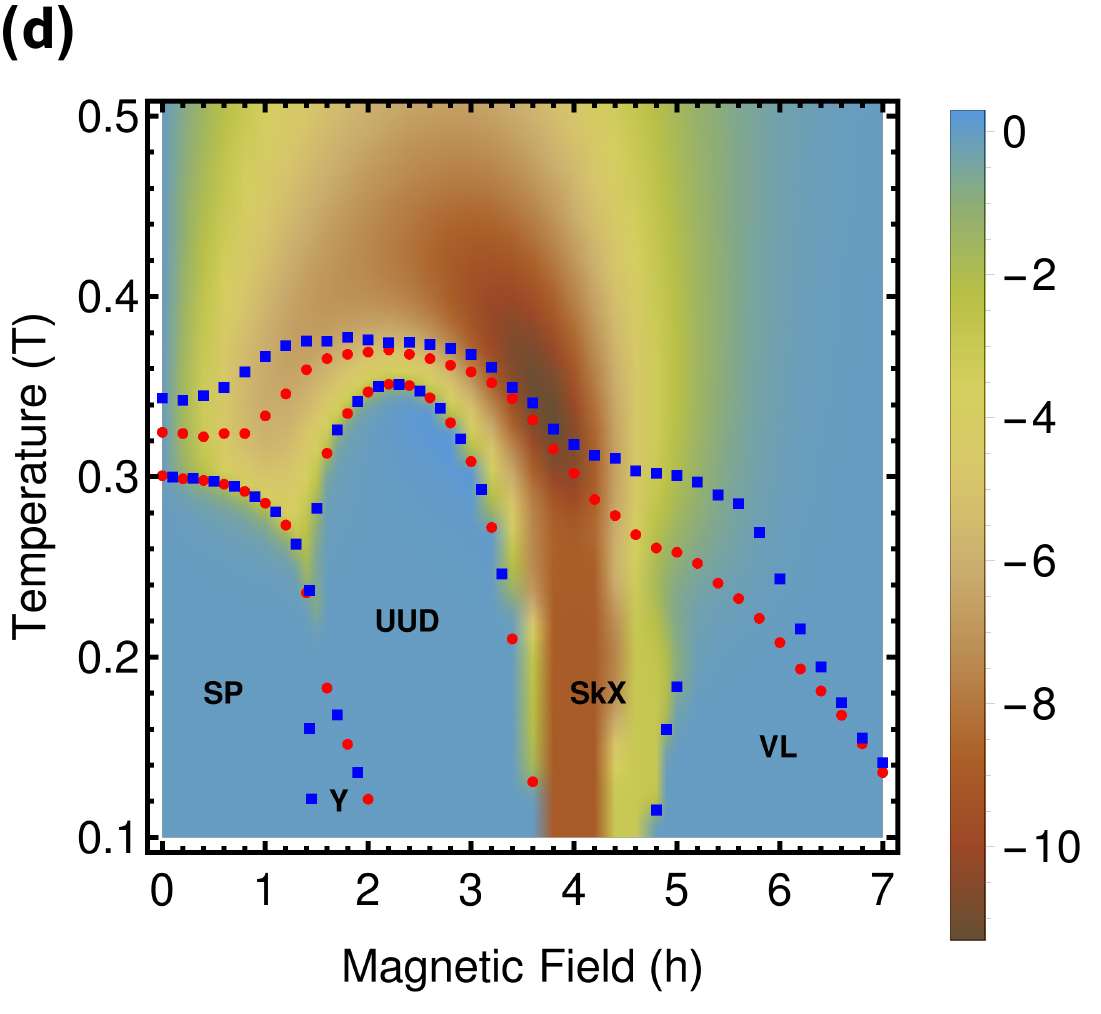}
\caption{Magnetic field--temperature (h--T) phase diagram of a triangular lattice AFM obtained by parallel tempering Monte Carlo simulations on a lattice of $75\times75$ atoms. The red circles correspond to peaks of the specific heat and the blue squares correspond to peaks of the topological susceptibility. (a) Phases SP, SkX, and VL are realized when $D_\parallel = 0.2$ and $K = -0.05$. (b) Phases SP, SkX, and VL are realized when $D_\parallel = 0.2$ and $K = 0$. (c) Phases SP, UUD, SkX, and VL are realized when $D_\parallel = 0.2$ and $K = 0.05$. (d) Phases SP, Y, UUD, SkX, and VL are realized when $D_\parallel = 0.2$ and $K = 0.1$. The color indicates the total topological charge in a region of $75\times75$ spins.}
 \label{fig6}
\end{figure*} 

The results of our calculations are shown in Fig.~\ref{fig6}. In the absence of DMI, our results agree with the known results~\cite{Kawamura1985,Gvozdikova,Seabra} (not shown). In a small magnetic field, a 120$\degree$ three-sublattice state is canted, forming a Y state with one spin aligned by the magnetic field and two others forming an angle with the opposite direction. At fields $h>3$, we observe a coplanar V canted state where two spins are aligned with each other and form V with the third spin. Note that at low temperatures, the degeneracy between the collinear and coplanar states is lifted by the order-from-disorder effects~\cite{Villain1980}. At $T=0$, $h=3$ and in the intermediate region, the system orders into UUD state with two spins aligned along the magnetic field and the third spin aligned in the opposite direction (see Fig.~1 in Ref.~\cite{Gvozdikova} for representations of Y, V, and UUD states).  

In the presence of DMI, we identify regions in the phase diagram corresponding to the spiral, antiferromagnetic SkX, and vortical-like (VL) states (see Figs.~\ref{fig6}(a), 8(b), 8(c), and 8(d)). Our results mostly agree with Ref.~\cite{Rosales} except for the presence of VL phase. According to our results, antiferromagnetic SkX phase appears even for relatively small strength of DMI, $D_\parallel=0.2$, as can be seen in Fig.~\ref{fig6}. At low temperatures and low magnetic fields the system prefers the spiral state as a natural extension of the behavior at $T = 0$ in Fig.~\ref{fig3}.
Increasing the magnetic field or temperature drives system into three-sublattice SkX phase.
In Fig.~\ref{fig5}(b), we show the spin structure factor for SkX phase realizable for $K= 0$, $D_{\parallel} = 0.2$, $h=3.1$, and $T=0.02$.
A snapshot of spin texture corresponding to SkX phase is shown in Fig.~\ref{fig8}. 
The SP-SkX transition is identified by peaks in the specific heat and the topological susceptibility. The SkX-VL transition is identified by peaks in the topological susceptibility (see Fig.~\ref{fig6}). The topological charge can also be used to identify the transitions at lower temperatures as can be seen from color representation of the topological charge in Fig.~\ref{fig6}. The exact location of the transition to paramagnetic (P) phase is harder to identify as peaks for the specific heat and the topological susceptibility do not coincide, and the average topological charge is nonzero in P phase. This behavior has been reported recently for realizations of SkX in ultrathin films Pd/Fe/Ir(111)~\cite{Bttcher2018}. 
Due to the presence of a well defined temperature-driven phase transition from a spiral phase, accompanied by a sharp rise in topological charge, we predict that SkX phase will persist even up to very low magnetic fields, where $h=0.1$ is the lowest field considered in Fig.~\ref{fig6}. As expected, the topological charge vanishes at $h=0$.

We further explore the effect of the easy axis ($K>0$) and easy plane ($K<0$) magnetic anisotropies. In Fig.~\ref{fig6}(a), we plot the phase diagram for $K=-0.05$ which at $h=0$, $T=0$ corresponds to the spiral phase in Fig.~\ref{fig3}. We identify regions corresponding to the SP, SkX, and VL phases, with SkX occupying a larger portion of the phase diagram compared to Fig.~\ref{fig6}(b) corresponding to $K=0$. In Fig.~\ref{fig6}(c), we plot the phase diagram for $K=0.05$. We identify regions corresponding to the SP, SkX, and VL phases; however, the region of SkX phase is shrunk due to appearance of additional UUD region. The UUD region becomes larger as we increase the easy axis anisotropy to $K=0.1$, as can be seen in Fig.~\ref{fig6}(d). In addition, for $K=0.1$ we identify a pocket corresponding to Y phase. Overall, the presence of easy axis magnetic anisotropy is not beneficial for realization of the SkX phase. On the other hand, the easy plane magnetic anisotropy can be beneficial for driving a spiral phase into skyrmion phase by applying magnetic field.

Of the systems listed in Table~\ref{table}, $\XMS{Fe}$ and $\XWSe{Fe}$ seem to be the most promising candidates for realizing the SkX phase thanks to the presence of strong DMI. However, large magnetic anisotropy in $\XWSe{Fe}$ will tend to suppress this phase, according to the phase diagrams in Fig.~\ref{fig6}. By exploiting the high-temperature SkX regions in Fig.~\ref{fig6}, one can apply magnetic fields that are much smaller compared to the exchange scale given by $J/\mu_B$. For $\XMS{Fe}$ this field could be on the order of 20 T. In order to obtain skyrmions in a three-sublattice antiferromagnet, it is desirable to find materials with weaker exchange interaction and smaller magnetic anisotropy. However, a weak easy plane magnetic anisotropy can be beneficial for stabilizing the SkX phase.

\section{Conclusion}\label{Sec:Conclusion}

We have studied the phase diagram of triangular AFM in the presence of DMI and magnetic anisotropy. Our results based on parallel tempering Monte Carlo simulations suggest feasibility of realizing spiral and skyrmion phases in triangular AFMs. The spiral phase appears at low temperatures and weak magnetic fields due to the presence of DMI. By adding magnetic field and by raising the temperature, one can drive the system into the skyrmion lattice phase.
Our Monte Carlo simulations indicate that both spirals and skyrmion lattices can appear even for relatively weak DMI, but their stability will be hindered by magnetic anisotropy. However, a weak easy plane magnetic anisotropy can be beneficial for stabilizing the skyrmion phase. We observe that skyrmion lattices can be stabilized below the threshold $D_\parallel=0.2$ suggested in Refs~\cite{PhysRevB.96.024404,Mohylna2020}.

As a realization of triangular AFMs, we have considered TMD-based antiferromagnetic triangular lattices. We have used first-principles computations to determine the parameters of magnetic interactions in monolayer TMDs ($\MS$,$\WS$, and $\WSe$) onto which 4 different transition metals (Co, Mn, Cr, and Fe) are adsorbed. We have shown that there are preferred adsorption sites, according to the strength of the Hubbard-U potential, and that all the compounds are antiferromagnets with nonzero uniaxial anisotropy and DMI.
According to our phase diagrams, at low temperatures the ground state of $\XMS{Cr}$, $\XMS{Fe}$, and $\XWSe{Fe}$ is a three-sublattice spiral. By further lowering anisotropy using materials engineering, TMD-based antiferromagnetic triangular lattices, such as $\XWSe{Fe}$, can potentially exhibit skyrmion lattices. On the other hand, we predict that $\XMS{Fe}$ can potentially host a three-sublattice skyrmion crystal in the presence of experimentally feasible magnetic fields.

\begin{acknowledgments}
AAK and AR were supported by the U.S. Department of Energy, Office of Science, Basic Energy Sciences, under Award No. DE-SC0021019. WF, PHC, and KDB were supported by the National Science Foundation (NSF) through the Nebraska Materials Research Science and Engineering Center (MRSEC) (grant no. DMR-1420645) and grant No. DMR-1916275. This research used resources of the National Energy Research Scientific Computing Center (NERSC), a U.S. Department of Energy Office of Science User Facility operated under Contract No. DE-AC02-05CH11231.
Part of this work was also completed utilizing the Holland Computing Center of the University of Nebraska, which receives support from the Nebraska Research Initiative. 
\end{acknowledgments}
\bigskip

\section*{Appendix A: Fourier transformed Hamiltonian for energy minimization}\label{App}

The elements of the Fourier transform of the Hamiltonian in Eq.~\eqref{Hamq} are given by the following matrix:
\begin{equation}
H_{q}=\begin{pmatrix}
A_{1} & \text{m}_{1,2}(\mathbf{k})& \text{m}_{1,3}(\mathbf{k})\\
\text{m}^{*}_{1,2}(\mathbf{k}) & A_{2} & \text{m}_{2,3}(\mathbf{k})\\
\text{m}^{*}_{1,3}(\mathbf{k}) & \text{m}^{*}_{2,3}(\mathbf{k}) & A_{3}\\
\end{pmatrix}.
\end{equation}
The $A_{\alpha}=\text{diag}(0,0,-\text{K})$ is the single-ion anisotropy for sublattice atom $\alpha$ and the submatrix $\text{m}_{\alpha,\beta}$ denotes the interaction between neighboring spins with sublattice indices $\alpha$, $\beta=\{1,2,3\}$. Using $k$ and $l$ as the Cartesian coordinates indices, these interaction submatrices read:
\begin{widetext}
\begin{equation}
m^{k l}_{1,2}(\mathbf{k})=(1+e^{i \mathbf{k}\cdot(\mathbf{e}_{1}+\mathbf{e}_{2})}+e^{i \mathbf{k}\cdot(2\mathbf{e}_{1}-\mathbf{e}_{2})})\delta^{k l}+ D_\parallel\sum_{n}(\mathbf{e}_{1}+\mathbf{e}_{2}e^{i \mathbf{k}\cdot(\mathbf{e}_{1}+\mathbf{e}_{2})}+ \mathbf{e}_{3}e^{i \mathbf{k}\cdot(2\mathbf{e}_{1}-\mathbf{e}_{2})})^{n}\varepsilon^{k l n};
\end{equation}
\begin{equation}
 m^{k l}_{1,3}(\mathbf{k})=(1+e^{i \mathbf{k}\cdot(\mathbf{e}_{1}+\mathbf{e}_{2})}+e^{i \mathbf{k}\cdot(-\mathbf{e}_{1}+2\mathbf{e}_{2})})\delta^{k l}+ D_\parallel\sum_{n}(\mathbf{e}_{2}+\mathbf{e}_{1}e^{i \mathbf{k}\cdot(\mathbf{e}_{1}+\mathbf{e}_{2})}+ \mathbf{e}_{3}e^{i \mathbf{k}\cdot(-\mathbf{e}_{1}+2\mathbf{e}_{2})})^{n}\varepsilon^{k l n};
\end{equation}
\begin{equation}
m^{k l}_{2,3}(\mathbf{k})=(1+e^{i \mathbf{k}\cdot(\mathbf{e}_{1}-2\mathbf{e}_{2})}+e^{i \mathbf{k}\cdot(2\mathbf{e}_{1}-\mathbf{e}_{2})})\delta^{k l}+ D_\parallel\sum_{n}(\mathbf{e}_{3}+\mathbf{e}_{2}e^{i \mathbf{k}\cdot(\mathbf{e}_{1}-2\mathbf{e}_{2})}+ \mathbf{e}_{1}e^{i \mathbf{k}\cdot(2\mathbf{e}_{1}-\mathbf{e}_{2})})^{n}\varepsilon^{k l n},
\end{equation}
\end{widetext}
with $D_\parallel$ being the strength of the DMI in reduced units, $\delta^{kl}$ being the Kronecker delta symbol, $\mathbf{e}_{1}=\{1,0,0\}$, $\mathbf{e}_{2}=\{\frac{1}{2},\frac{\sqrt{3}}{2},0\}$ and $\mathbf{e}_{3}=\mathbf{e}_{1}-\mathbf{e}_{2}$, and $\varepsilon^{k l n}$ being the Levi-Civita tensor.
\section*{Appendix B: Feedback-optimized parallel tempering Monte Carlo}
The parallel tempering Monte Carlo algorithm simulates a set of noninteracting replicas at temperatures ${T_1, \dots}, T_M$ where $M$ is the number of replicas~\cite{Hukushima1996,PhysRevLett.57.2607,Marinari1992,Lyubartsev1992}. After performing a fixed number of sweeps (typically a few), the algorithm suggests a swap of replicas at neighbouring temperatures, $T_i$ and $T_{i+1}$, accepting the swap with probability,
\begin{equation}
    p(E_i,T_i \rightarrow E_{i+1},T_{i+1})=\min \{1,\exp(\Delta \beta \Delta E)\},
\end{equation}
where we introduce the difference between inverse temperatures, $\Delta \beta=1/( T_{i+1})-1/(T_{i})$, and the difference in energy of the two replicas, $\Delta E=E_{i+1}-E_i$. The algorithm leads to diffusion of replicas in temperature space. Visits of the high temperature regions, where relaxation happens faster, facilitate relaxation at lower temperatures. Thus, the system can efficiently relax in the presence of complicated energy landscapes. In this process, it is important to maximize the number of round trips between the lowest and the highest temperature, $T_1$ and $T_M$. The number of round trips will strongly depend on the choice of temperature set, ${T_1, \dots}, T_M$. A temperature set can be chosen according to a geometric progression. This, however, will result in suppressed exchanges between temperatures at a phase transition where specific heat diverges. 

\begin{figure}
\includegraphics[width=0.47\columnwidth]{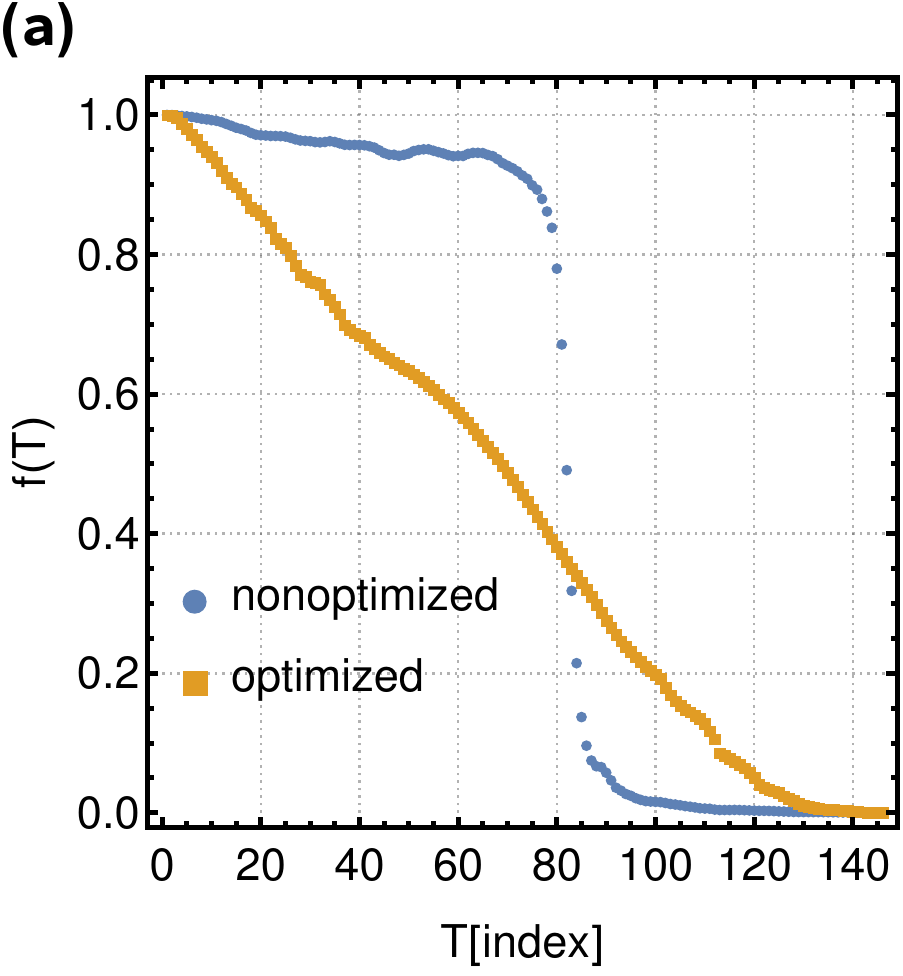}
\includegraphics[width=0.49\columnwidth]{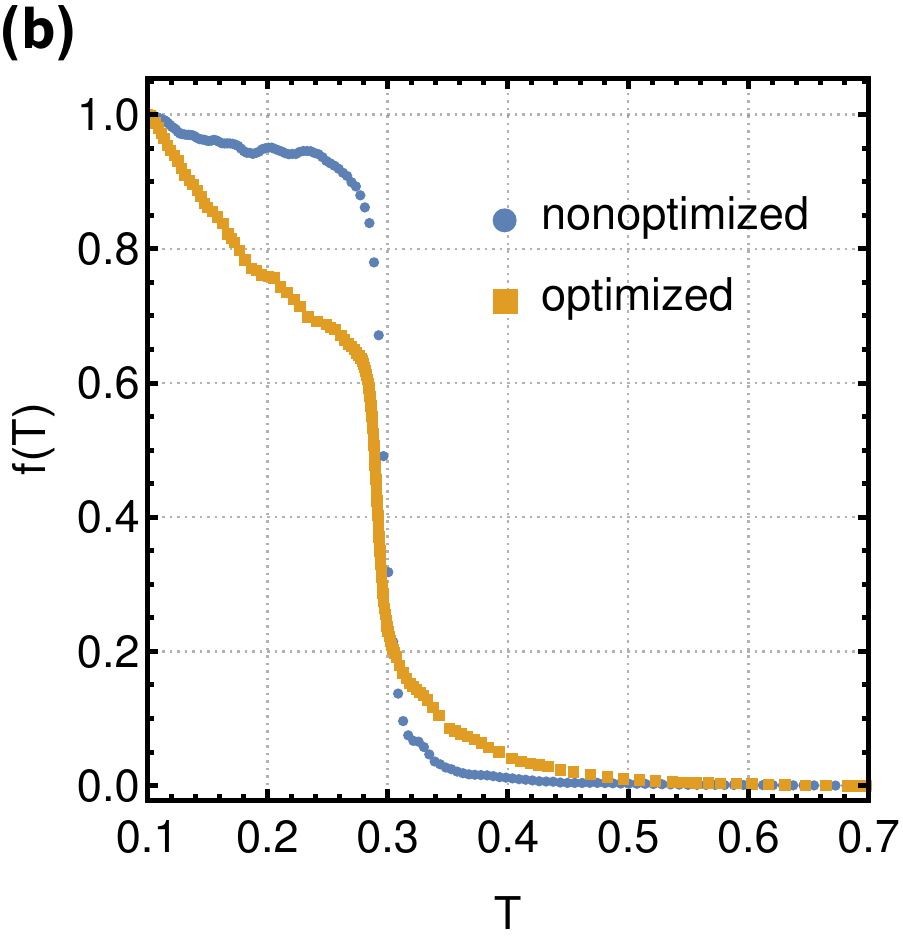}
\caption{Fraction of temperatures $f(T)$ moving from the lowest to the highest temperature as a function of the temperature index in (a), and as a function of temperature in (b). The calculation corresponds to parameters: $D_\parallel= 0.2$, $K = 0.1$, and $h=0.8$.}
\label{appendix}
\end{figure}

In this work, we use approach suggested in Ref.~\cite{Katzgraber2006} where the temperature set is optimized by measuring the diffusion of replicas. In particular, a replica assumes index ``up" after visiting $T_1$ and a replica assumes index ``down" after visiting $T_M$.
The index is rewritten from ``up" to ``down" every time the ``up" replica visits $T_M$, and
the index is rewritten from ``down" to ``up" every time the ``down" replica visits $T_1$. 
After each sweep, we update quantities $n_\text{up}(T_i)$ and $n_\text{down}(T_i)$ for each $T_i$ by adding $1$ to $n_\text{up}(T_i)$ if the replica at $T_i$ has index ``up" and by adding $1$ to $n_\text{down}(T_i)$ if the replica at $T_i$ has index ``down". 
The quantity of interest, $f(T)$, evaluated for each temperature characterizes the fraction of replicas which have visited either $T_1$ or $T_M$: 
\begin{equation}
    f(T_i)=\frac{n_\text{up}(T_i)}{n_\text{up}(T_i)+n_\text{down}(T_i)},
\end{equation}
where $n_\text{up/down}(T_i)$ counts the total number of ``up"/``down" walkers throughout the whole Monte Carlo simulation for a temperature $T_i$. Using function $f(T_i)$ recalculated for each temperature set, we iteratively maximize the current of replicas in temperature space~\cite{Katzgraber2006}. For an optimal temperature set, the function $f(T_i)$ will decrease at a constant rate as a function of the temperature index, as seen in Fig.~\ref{appendix}(a). 
\begin{figure}
 \includegraphics[width=0.9\linewidth]{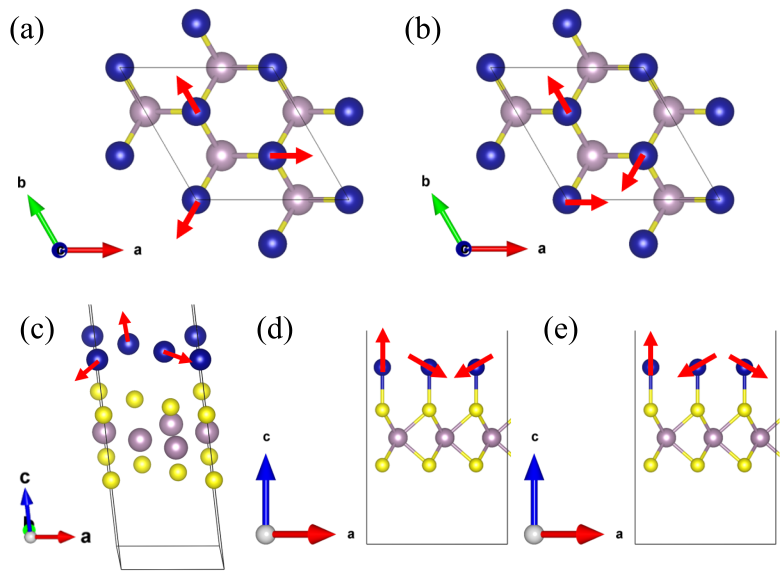}
 \caption{Spin configurations for extracting parameters in the model Hamiltonian: (a) and (b) in-plane coplanar 120\degree\ spin configurations with an opposite chirality. (c) an out-of-plane coplanar 120\degree\ spin configuration. (d) and (e) coplanar 120\degree\ spin configurations in a $3\times1$ unit cell.}
 \label{appendix2}
\end{figure} 
In Fig.~\ref{appendix}, we give an example of $f(T_i)$ calculated before (a geometric progression) and after the optimization procedure. To increase the temperature diffusion, temperatures in a temperature set are shifted toward the region of a phase transition, as can be seen in Fig.~\ref{appendix}(b). 

\section*{Appendix C: Extracting parameters from model Hamiltonian}
To extract parameters of Hamiltonian \eqref{Ham}, we perform several constrained DFT calculations with specified non-collinear spin configurations. To account for the $120\degree$ in-plane antiferromagnetic ordering, we use a $\sqrt{3}\times\sqrt{3}$ unit cell (three transition atoms per unit cell) to perform the calculations. The nearest-neighbor exchange parameter $J$ is obtained without spin-orbit interaction, which also results in vanishing anisotropy and DMI. We tilt the spins out of plane by several small angles. The energy difference with respect to the in-plane spin configuration as a function of the tilting angle $\theta$ is $\Delta E\approx\frac{27}{2}J\theta^2$ which allows us to determine the exchange parameter by fitting with the energy as a function of the tilting angle $\theta$. The single-ion anisotropy $K$, and in-plane and out-of-plane DMI components can be determined by taking the energy difference between suitably specified spin configurations. For out-of-plane DMI component, we use two in-plane coplanar 120\degree\ spin configurations with opposite chirality as shown in Figs.~\ref{appendix2}(a) and (b). The energy difference between these two spin configurations is $\Delta E=9\sqrt{3}\Tilde{\text{D}}_{\perp}$. For single-ion anisotropy $K$, we use one in-plane and one out-of-plane coplanar 120\degree\ spin configurations as shown in Figs.~\ref{appendix2}(a) and (c). The energy difference between these two spin configurations is $\Delta E=\frac{9}{2}\sqrt{3}\Tilde{\text{D}}_{\perp}+\frac{3}{2}K$. For in-plane DMI component, we notice that it is not possible to obtain it in a $\sqrt{3}\times\sqrt{3}$ unit cell as the contribution to the total energy from in-plane DMI component is canceled out by symmetry regardless of how we specify the spin configurations. To break this symmetry, a $3\times1$ unit cell (three transition atoms per unit cell) is used for the calculation of the in-plane DMI component. We use two spin configurations with opposite chirality as shown in Figs.~\ref{appendix2}(d) and (e). The energy difference between these two spin configurations is $\Delta E=\frac{9}{2}\sqrt{3}\Tilde{\text{D}}_{\parallel}$.

\bibliography{Bib1}
\end{document}